\documentclass[aps,pra,amsmath,amssymb,11pt,final,tightenlines,twoside,onecolumn,nofloats,nofootinbib,superscriptaddress,
showkeys,showkeywords]{revtex4-2}

\usepackage[T2A]{fontenc}
\usepackage[utf8x]{inputenc}
\usepackage[russian,english]{babel}
\usepackage{graphicx}

\begin{document}

\selectlanguage{english}

\noindent {\it ASTRONOMY REPORTS, 2025, volume , No. }
\bigskip\bigskip  \hrule\smallskip\hrule
\vspace{35mm}


\title{STATISTIC OF STARS WITH POOR ASTROMETRIC SOLUTIONS OF GAIA DR3 IN OPEN STAR CLUSTERS}

\author{\bf\copyright~2025 г. \firstname{D.~I.}~\surname{Tagaev}}
\email{dima.tagaev@list.ru}
\affiliation{Ural Federal University, Lenin str. 51, Ekaterinburg, 620000 Russia}

\author{\bf\firstname{A.~F.}~\surname{Seleznev}}
\email{anton.seleznev@urfu.ru}
\affiliation{Ural Federal University, Lenin str. 51, Ekaterinburg, 620000 Russia}

\begin{abstract}
\vspace{3mm}
\received{12.06.24}
\revised{12.06.24}
\accepted{12.06.24} 
\vspace{3mm}

We performed counts of stars with poor astrometric solutions of Gaia DR3 in the regions of open star clusters: NGC 188, NGC 1039, NGC 2287, NGC 2301, NGC 2360, NGC 2420, NGC 2527, NGC 2548, NGC 2682 (M 67), NGC 3114, NGC 3766, NGC 5460, NGC 6649.
The selection of the possible cluster members is based on the Gaia photometry using the Hess diagram.
We look for stars that fall within the region of the Hess diagram plotted from probable cluster members based on Hunt\&Reffert data.
We take stars with two-parameter solutions, with the parameter RUWE>1.4, as well as with large relative parallax errors that fall within the Hess diagram region for probable cluster members.
The radii of clusters based on stars with poor astrometric solutions and the number of such possible cluster members were estimated.
The number of stars with poor astrometric solutions relative to the number of stars from the Hunt\&Reffert sample $N_\textrm{bad}/N$ varies very widely with a median average of approximately 30\%.
This means that when one selects probable cluster members based on precise astrometric data from Gaia DR3, an average of about 23\% of cluster members may be lost.
Among the lost stars there may be a significant number of unresolved binary and multiple systems.
We investigated the dependence of the relative number of stars with poor astrometric solutions on the galactic latitude and on the average number density of stars.
The brightness functions with and without stars with poor solutions differ significantly in the region of faint stars ($G\in[14,18]$ mag) for clusters with a relative number of possible cluster members with poor astrometric solutions of $N_\textrm{bad}/N\gtrsim0.15$.

\end{abstract}

\maketitle

\section{INTRODUCTION}

In the previous work \cite{Tagaev} we successfully tested the method of searching for possible members of the cluster with poor astrometric solutions of Gaia DR3.
We showed that with the traditional methods of extracting members of open clusters from accurate astrometric data of Gaia DR3, almost 50\% of the members of the cluster NGC 3532 \cite{Tagaev} could be lost.
It is of interest to consider a larger number of clusters in order to obtain statistics of stars with poor astrometric solutions in open clusters.

This study is relevant because the samples of probable cluster members of \cite{H&R2024} and \cite{Cantat-Gaudin+2020} cannot be considered as complete, although they are the basis for studying the kinematics and dynamics of clusters.
For example, any sample will be limited by the limiting stellar magnitude.
However, the main problem is that when one obtains a sample, various restrictions are imposed on astrometric parameters.
Selection occurs at the base of parallaxes and proper motions, although about 19\% of all stars in Gaia DR3 \cite{GaiaDR3} have only two-parameter solutions.
In addition, the sample usually is limited by the parameter RUWE<1.4 (renormalized unit weight error).
Authors often use a limitation on the relative parallax error, see for example \cite{El-Depsey+2023,Koc}.
The purpose of such limitations is clear.
The authors try to obtain reliable samples of probable cluster members, but in doing so the completeness of these samples suffers.

Nevertheless, completeness of samples is necessary for a number of tasks, for example, to obtain the brightness function and mass spectrum.
Thus, authors of \cite{Tagaev} showed a significant difference between the brightness functions constructed only for cluster members selected by parallax and proper motions, and taking into account possible members with poor astrometric solutions, in the range of $G\in[14,18]$ mag.
Also, to analyze the cluster dynamics and model the evolution of the cluster, one needs to know the total mass of the cluster and the number of its stars.

Poor astrometric solutions arise due to several reasons.
One of the main reasons is the use of a single star motion model to determine the astrometric parameters in Gaia DR3.
This model incorrectly describes the motion of more complex objects, such as binary and multiple stars.
The two-parameter solutions, solutions with large values of the relative parallax error $\sigma_\varpi/\varpi$ and RUWE \cite{Lindegren+2021} emerge due to this point.
Therefore, among the stars not included in the sample of the probable cluster members, there may be a large number of unresolved binary and multiple systems.

Parallax shifts for bright stars ($G<13$) \cite{C-G_B,Ding} could be among other reasons.
Also, the systematic errors of parallax and proper motions of the photocenter of reference eclipsing binary stars \cite{Stassun} can make their contribution, which in turn affects the RUWE parameter.

The aim of our work is to determine, at the base of the Hunt\&Reffert \cite{H&R2024} sample, how many stars with poor astrometric solutions of Gaia DR3 may be members of open clusters and how these stars may influence the overall cluster brightness function.
In Section II we select stars with poor astrometric solutions in a wide neighborhood of the cluster that fall within the region occupied by probable cluster members from \cite{H&R2024}.
In Section III we estimate the number of possible cluster members with poor astrometric solutions.
In Section IV we study the influence of such stars on the cluster brightness function.
Section V is devoted to a discussion of the results of the work.

\section{SELECTION OF THE POSSIBLE CLUSTER MEMBERS WITH POOR ASTROMETRIC SOLUTIONS OF GAIA DR3 USING THE HESS DIAGRAM}

The Hess diagram is a two-dimensional probability density map plotted in the coordinates of stellar magnitude and color index.
To construct it, we use the Kernel Density Estimator (KDE) method \cite{Silverman1986}.
The two-dimensional KDE method with a biquadratic kernel is used \cite{Silverman1986} to construct the Hess diagram:

\begin{equation}
\label{kernel}
    K(x)= \left\{
            \begin{array}{ll}
\frac{\displaystyle 3}{\displaystyle \pi h^2}\left(1-\frac{\displaystyle x^2+y^2}{\displaystyle h^2}\right)^2 & ,\; x^2+y^2\leq h^2 \\
0                & ,\; x^2+y^2 > h^2  \\
            \end{array}
    \right.
\end{equation}

\noindent where $x$ and $y$ are the distance from the data point to the grid node at which the function value is determined, $h$ is the kernel half-width.
We calculate values of the density in the nodes of a rectangular grid.
To go from density values to probability values, one should divide all the density values at the nodes by the total number of points used to construct the Hess diagram.
We use the biquadratic kernel because it has a finite size, which allows for significant savings in computation time.
In addition, unlike the quadratic kernel, the biquadratic kernel gives a smoother estimate.
The kernel half-width for each cluster is selected individually, in accordance with the appearance of the three-dimensional Hess diagram \cite{Tagaev}, in order to obtain a probability density function without sharp changes.

\begin{figure}[htbp]
	\includegraphics[width=1.0\textwidth]{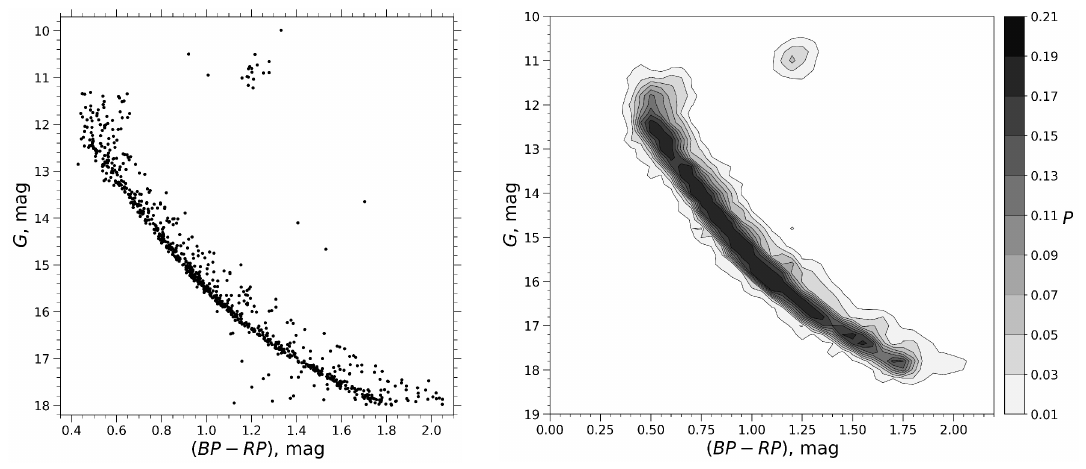}
	\caption{CMD of the cluster NGC 2360 (left). Hess diagram for NGC 2360 (right), plotted with a kernel half-width of 0.4 mag; grayscale shows probability values. }
	\label{CMD}
\end{figure}

\begin{figure}[htbp]
	\includegraphics[width=1.0\textwidth]{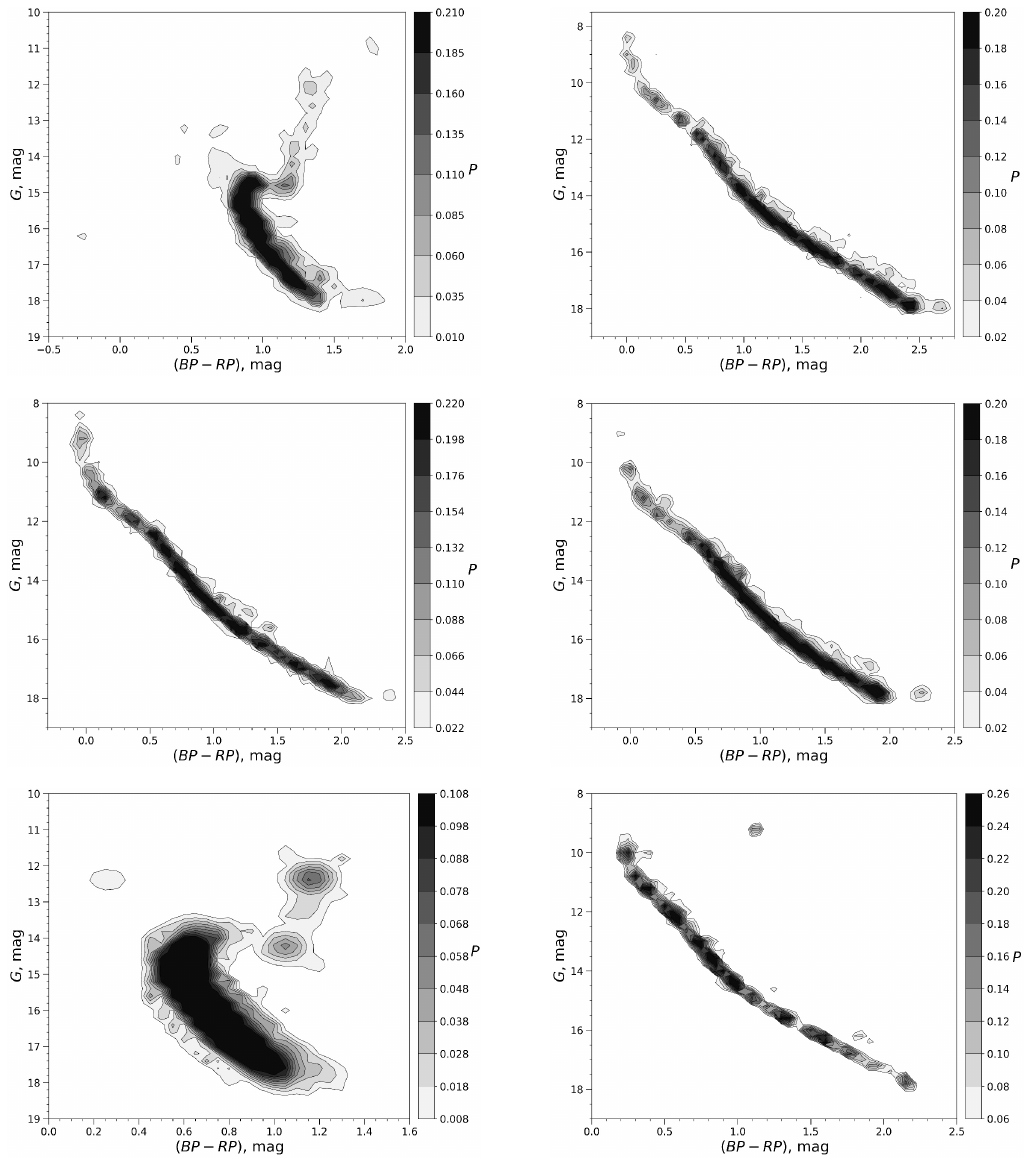}
	\caption{Hess diagrams for clusters, from left to right, top to bottom: NGC 188, NGC 1093, NGC 2287, NGC 2301, NGC 2420, NGC 2527. The Hess diagram for NGC 2420 was plotted with a kernel half-width of $h=0.6$ mag, for the others a kernel half-width of $h=0.3$ mag was used. }
	\label{Hess_1}
\end{figure}

\begin{figure}[htbp]
	\includegraphics[width=1.0\textwidth]{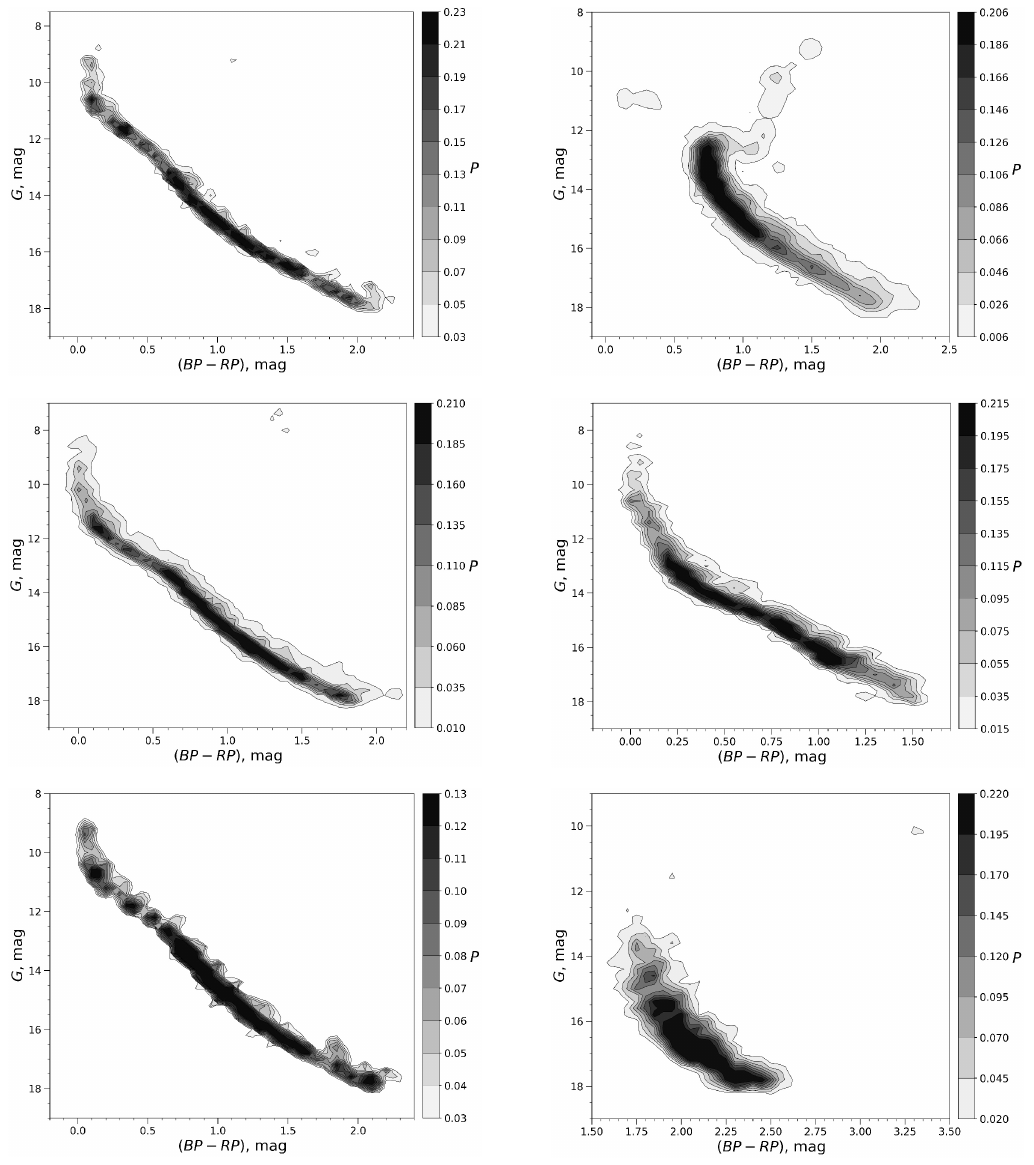}
	\caption{Hess diagrams for clusters, from left to right, top to bottom: NGC 2548, NGC 2682, NGC 3114, NGC 3766, NGC 5460, NGC 6649. The Hess diagram for NGC 2682 was plotted with a kernel half-width of $h=0.5$ mag, for NGC 5460 --- $h=0.4$ mag, for the rest a kernel half-width of $h=0.3$ mag was used. }
	\label{Hess_2}
\end{figure}

In this study, we restrict ourselves to the limiting stellar magnitude $G=18$, since for fainter stars the errors in the astrometric parameters of Gaia DR3 increase sharply.
We plot Hess diagrams for clusters using stars from the Hunt\&Reffert \cite{H&R2024} sample with a membership probability greater than 50\%.
As an example, Fig. \ref{CMD} (left panel) shows a color-magnitude diagram (CMD) of such stars for the cluster NGC 2360.
The resulting Hess diagram with an optimal kernel half-width of 0.4 mag for this cluster is shown in Fig. \ref{CMD} at the right panel.
The grayscale shows the range of probability values.
The figure shows a shortened probability range to better show areas of the chart with low probability values.
The Hess diagrams of the remaining clusters are shown in Fig. \ref{Hess_1} and Fig. \ref{Hess_2} with the kernel half-widths used indicated.

\begin{figure}[htbp]
	\includegraphics[width=0.5\textwidth]{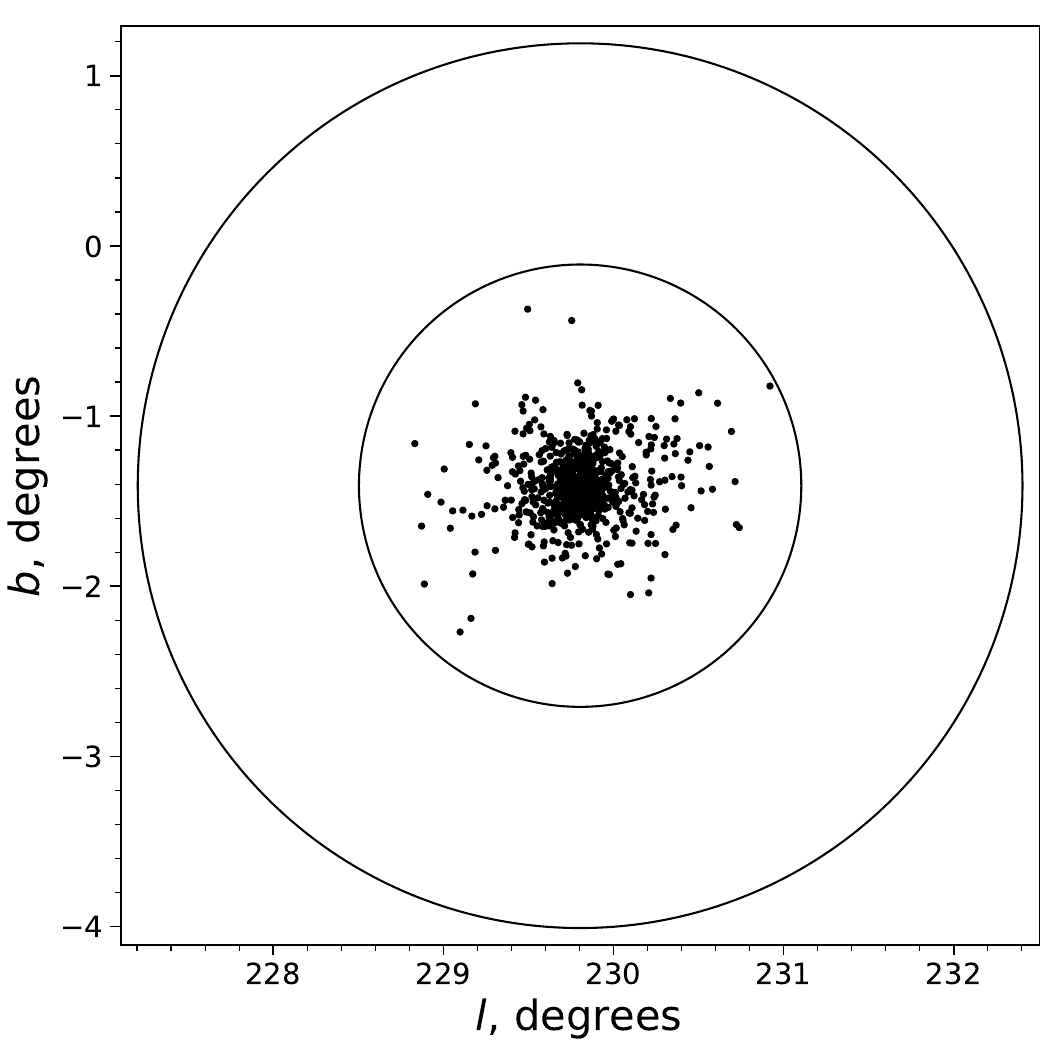}
	\caption{Map of the distribution of stars of the Hunt\&Reffert sample \cite{H&R2024} for NGC 2360 cluster, with the cluster region as the smaller circle and the search region as the larger circle. }
	\label{area}
\end{figure}

To select probable cluster members with poor astrometric solutions of Gaia DR3, we need first to determine from the known sample \cite{H&R2024} the region of the sky where the cluster members lie.
Fig. \ref{area} presents the members of the NGC 2360 cluster from the \cite{H&R2024} sample with $G\leq18$ mag and all membership probabilities.
The smaller circle is the cluster region with the center at $l_\textrm{c}=229.804^\circ$, $b_\textrm{c}=-1.409^\circ$ \cite{SIMBAD} and the radius $R_\textrm{c}=1.3^\circ$ (chosen so that all probable cluster members are inside this circle).
The larger circle is the search area for possible cluster members with the same center and radius $R=2R_\textrm{c}$.
In the search area we select stars with $G\leq18$ mag (the star must have photometry data in three bands of Gaia D3), provided that the star is not included in the sample of \cite{H&R2024}.
In this area, we select stars with two-parameter solutions or stars with 5- or 6-parameter solutions, but having the parameter RUWE>1.4, and/or the relative parallax error $\delta_\varpi/\varpi>0.2$.
For these stars, we determine the probability of being a member of a cluster using the Hess diagram.
As a lower threshold probability value, we take the maximum value of the function (\ref{kernel}), which is approximately 10.6, divided by the number of stars used to construct the Hess diagram (individual for each cluster).
By the order of magnitude, the threshold value is approximately 1\%.
Thus, for the cluster NGC 2360, the number of stars used to construct the Hess diagram is 612.
Therefore, the probability threshold for this cluster is approximately 1.7\%.
This is necessary in order to exclude the areas of "lonely" stars on the Hess diagram, which could contain a large number of background stars.

Thus, we obtained samples of possible cluster members with poor astrometric solutions, which are located on the CMD in the same place as the probable cluster members from the Hunt\&Reffert \cite{H&R2024} sample.

\section{ESTIMATION OF THE NUMBER OF POSSIBLE MEMBERS OF OPEN CLUSTERS WITH POOR ASTROMETRIC SOLUTIONS}

\begin{figure}[htbp]
	\includegraphics[width=0.6\textwidth]{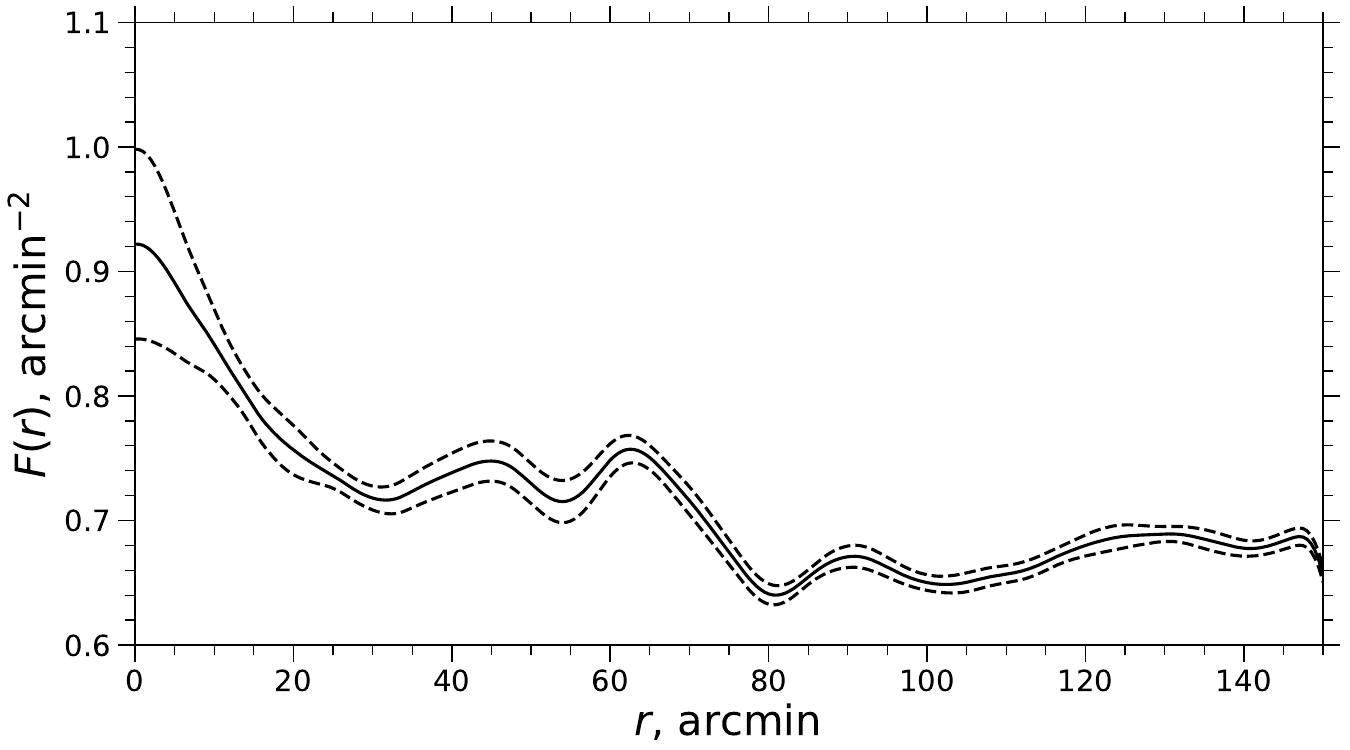}
	\caption{Radial profile of the surface density of the cluster NGC 2360, plotted by the KDE method \cite{Silverman1986} with a kernel half-width of 10 arcmin.
		The dashed line shows the confidence interval of $2\sigma$ width. }
	\label{Profile}
\end{figure}

\begin{figure}[htbp]
	\includegraphics[width=0.6\textwidth]{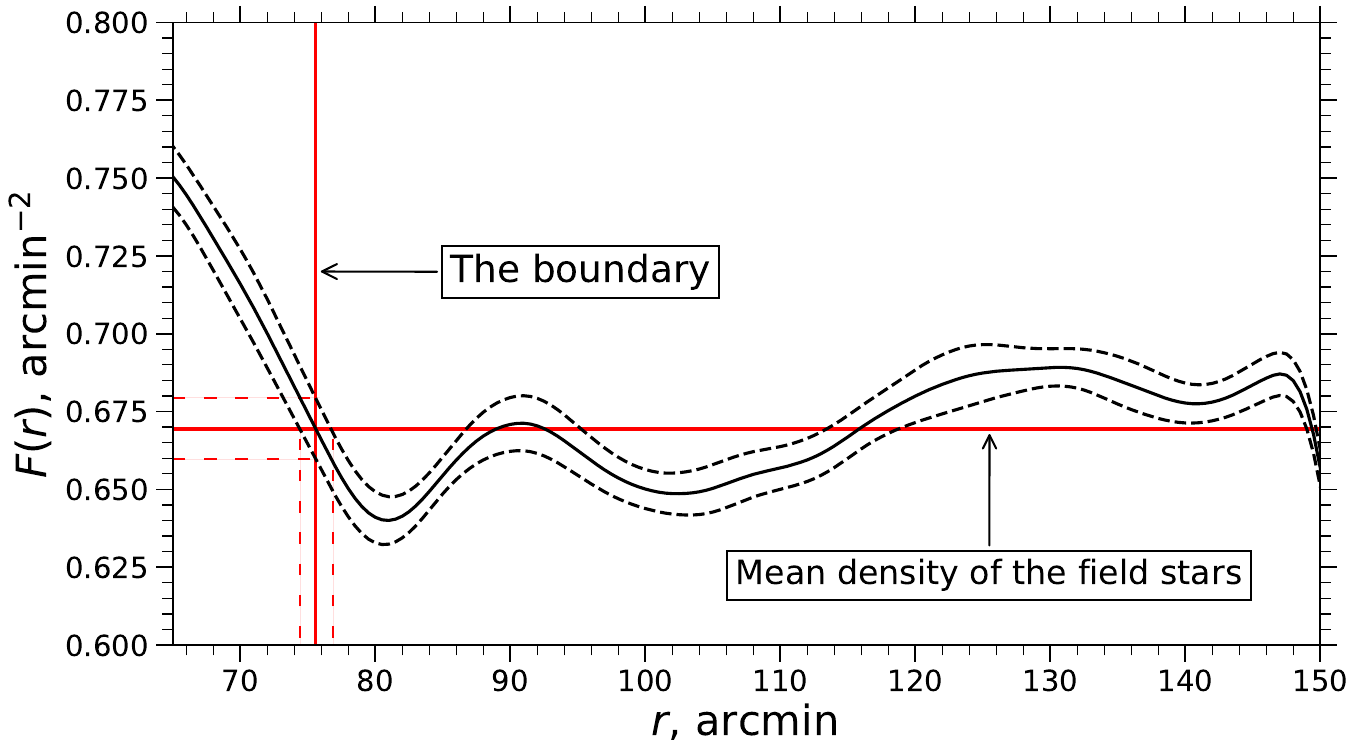}
	\caption{Determination of the radius of the cluster NGC 2360 and the average density of field stars by the method of \cite{Seleznev2016} based on stars with poor astrometric solutions of Gaia DR3.
		The solid black line is the density profile, the dotted black lines are the confidence interval of $2\sigma$ width. }
	\label{R}
\end{figure}

Our samples contain a large number of field stars.
To estimate the number of possible cluster members with poor solutions, we use the method of \cite{Seleznev2016}.
To do this, we plot a radial profile of the surface density \cite{Seleznev2016} for the stars we selected in Section II, using the KDE method with a biquadratic kernel.
The key point is the presence (or absence) of a concentration of stars with poor astrometric solutions towards the cluster center.
If there is a concentration towards the center, this means that among the stars with poor astrometric solutions there are possible cluster members.
In this case, the analysis of the radial profile allows us to estimate the cluster radius by the stars with poor astrometric solutions of Gaia DR3 and the number of such stars --- possible members of the cluster.
As an example, the figure \ref{Profile} shows the radial density profile of the cluster NGC 2360 with a confidence interval determined using the smoothed bootstrap method \cite{Seleznev2016}.
To plot the radial density profile, a kernel half-width of $h=10$ arcmin was used, which is optimal in this case if we follow the methodology of \cite{Merritt&Tremblay1994,Seleznev2016}.
The enlarged region of the radial profile in Figure \ref{R} shows the determination of the cluster radius and the average field star number density for stars with poor astrometric solutions (the method is described in more detail in the works \cite{Seleznev2016,Tagaev}).

In the case of the cluster NGC 2360, the cluster radius based on stars with poor astrometric solutions is $R_\textrm{b}=(75.6\pm1.2)$ arcmin and the average background density is $\rho_\textrm{b}=(0.67\pm0.01)$ $\textrm{(arcmin)}^{-2}$.
Table~\ref{table} lists the obtained cluster radii.
For the cluster NGC 188, we did not find the concentration of stars with poor astrometric solutions towards the center.
Consequently, it is impossible to determine the cluster radius in the case of NGC 188.

\begin{table}[ht]
	\caption{Number of possible cluster members with poor astrometric solutions of Gaia DR3}
	\label{table}
	\begin{tabular}{c|c|c|c|c}
		\hline
		Cluster  & $b$, deg    & $R_\textrm{b}$, arcmin & $N_\textrm{bad}/N$ & $\overline{F}$, $(\textrm{arcmin})^{-2}$ \\
		\hline
		NGC 2301 & 0.286 & $52.6\pm1.9$ & $0.43\pm0.08$ & {} \\ 
		NGC 2301 & 0.286 & $120\pm18$ & $2.7\pm0.4$ & 4.190 \\
		NGC 3766 & -0.037 & $11.8\pm0.5$ & $0.45\pm0.06$ & 16.501 \\ 
		NGC 6649 & -0.780 & $4.8\pm0.4$ & $0.050\pm0.015$ & 6.016 \\
		NGC 3532 & 1.375 & $210\pm3$ \cite{Tagaev} & $1.00\pm0.11$ \cite{Tagaev} & 13.410 \\
		NGC 2360 & -1.409 & $75.6\pm1.2$ & $1.42\pm0.24$ & 4.533 \\
		NGC 2527 & 1.871 & $12.9\pm2.6$ & $0.035\pm0.022$ & 6.461 \\
		NGC 3114 & -3.805 & $118\pm12$ & $3.9\pm0.5$ & 10.204 \\
		NGC 2287 & -10.432 & $56\pm21$ & $0.04\pm0.11$ & 2.830 \\
		NGC 5460 & 12.687 & $58.3\pm3.6$ & $0.6\pm0.4$ & 4.201 \\
		NGC 2548 & 15.390 & $281\pm9$ & $0.3\pm0.5$ & 1.558 \\
		NGC 1039 & -15.646 & $123\pm4$ & $0.13\pm0.12$ & 1.502 \\
		NGC 2420 & 19.640 & $18\pm5$ & $0.07\pm0.07$ & 1.060 \\
		NGC 188  & 22.374 & --- &  --- & --- \\
		NGC 2682 & 31.921 & $30.1\pm2.4$ & $0.030\pm0.008$ & 0.391 \\
		\hline
	\end{tabular}
	
	The first column contains the cluster name, the second column contains the galactic latitude $b$, the third column contains the cluster radius based on stars with poor astrometric solutions of Gaia DR3 $R_\textrm{b}$, the fourth column contains the relative number of possible cluster members with poor astrometric solutions of Gaia DR3 (the ratio of the number of stars with poor solutions to the number of all cluster members in the Hunt\&Reffert sample \cite{H&R2024} with $G\leq18$ mag and all membership probabilities) $N_\textrm{bad}/N$, and the fifth column contains the average surface density of all stars in the search area for each cluster $\overline{F}$.
\end{table}

To obtain the number of possible cluster members with poor astrometric solutions of Gaia DR3, we subtract the number of field stars with density $\rho_\textrm{b}$ from the number of stars in the cluster circle with radius $R_\textrm{b}$ \cite{Tagaev}.
Table~\ref{table} lists the relative numbers of possible cluster members with poor astrometric solutions $N_\textrm{bad}/N$ for all clusters.
$N_\textrm{bad}$ is the number of possible cluster members with poor astrometric solutions, $N$ is the number of probable cluster members with $G\leq18$ mag. and any membership probability values from the \cite{H&R2024} sample.
We give two options for the cluster NGC 2301 because it is difficult to select unambiguously a field region of the radial density profile (based on stars with poor astrometric solutions) for this cluster.
The median of the distribution of the relative number of possible members with poor astrometric solutions is approximately 0.3.
Thus, with the traditional method of identifying probable cluster members, one can lose 23\% of stars (on average) because they have poor astrometric solutions.

For some clusters $N_\textrm{bad}/N\geqslant1$, as in the case of NGC 3532 \cite{Tagaev}.
This means that the number of possible cluster members with poor astrometric solutions is equal to or greater than the number of stars in the Hunt\&Reffert sample \cite{H&R2024}.
The only possible explanation, in our opinion, is that in this region there are actually a lot of stars with poor astrometric solutions.
Therefore, a study like our one is necessary for each cluster, especially when it comes to obtaining the luminosity function and mass spectrum.

\begin{figure}[htbp]
	\includegraphics[width=0.495\textwidth]{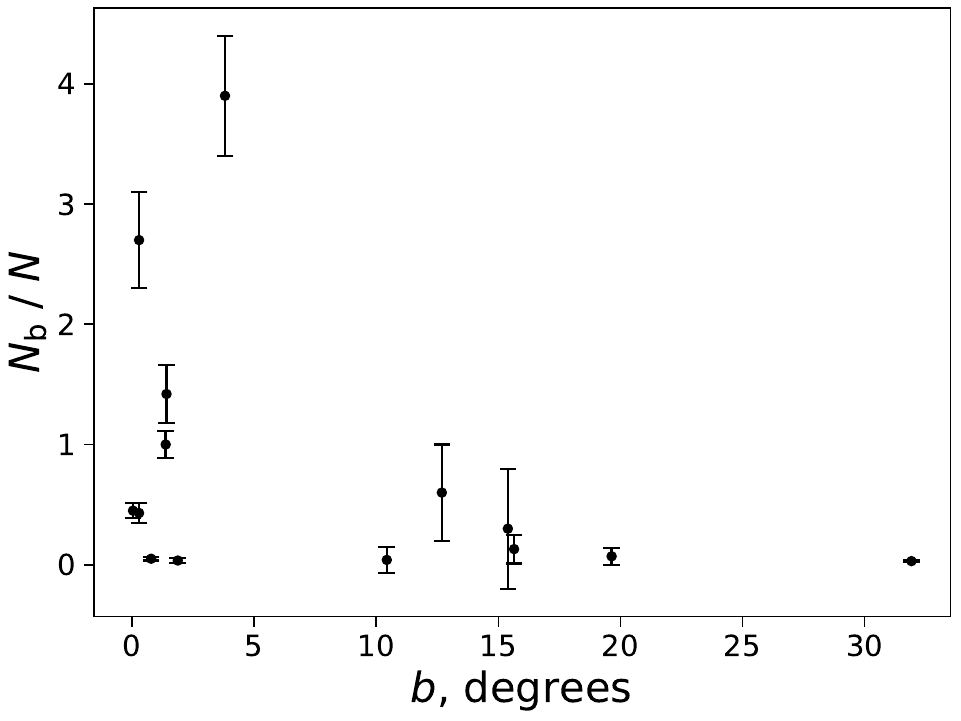}
	\includegraphics[width=0.495\textwidth]{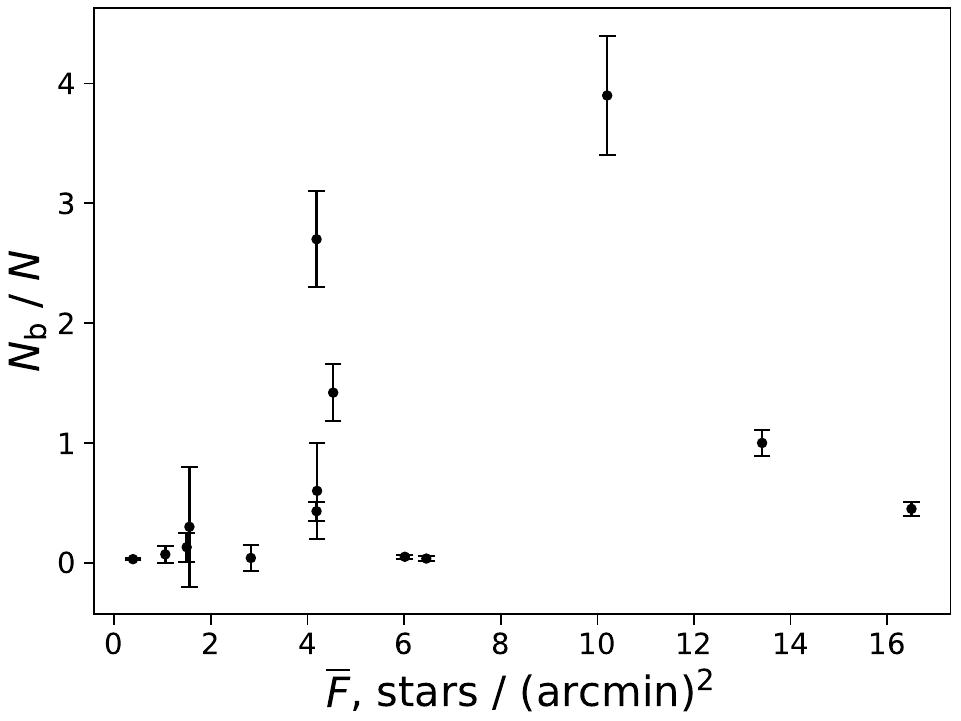}
	\caption{Distribution of the relative number of stars by galactic latitude $b$ (left panel).
		Distribution of the relative number of stars by the average surface density of all stars $\overline{F}$ (right panel). }
	\label{N}
\end{figure}

Figure \ref{N} shows the distributions of the relative number of stars with poor astrometric solutions by the galactic latitude $b$ (left panel) and by the average surface density of all stars in the search area $\overline{F}$ (right panel).
We determined the error bars in figure \ref{N} using the errors of the number of stars with poor astrometric solutions.
In the left panel of figure \ref{N} one can see that at small values of latitude $b$ there are clusters with a large value of the relative number of possible members with poor astrometric solutions.
As $b$ increases, the number of such clusters decreases significantly. 
The right panel of the figure \ref{N} shows an increase in the relative number of possible cluster members with poor astrometric solutions with increasing average density of all stars (the increase in the sense of the increase of the upper envelope line drawn over the points).
Thus, we can assume that one of the sources of poor astrometric solutions of Gaia DR3 is the high density of stars in the cluster region (galactic latitude is obviously related to the overall density of stars).
Note that the number of clusters in our sample is not sufficient to draw definitive conclusions.

\section{INFLUENCE OF STARS WITH POOR ASTROMETRIC SOLUTIONS ON THE CLUSTER BRIGHTNESS FUNCTION}

\begin{figure}[htbp]
	\includegraphics[width=0.6\textwidth]{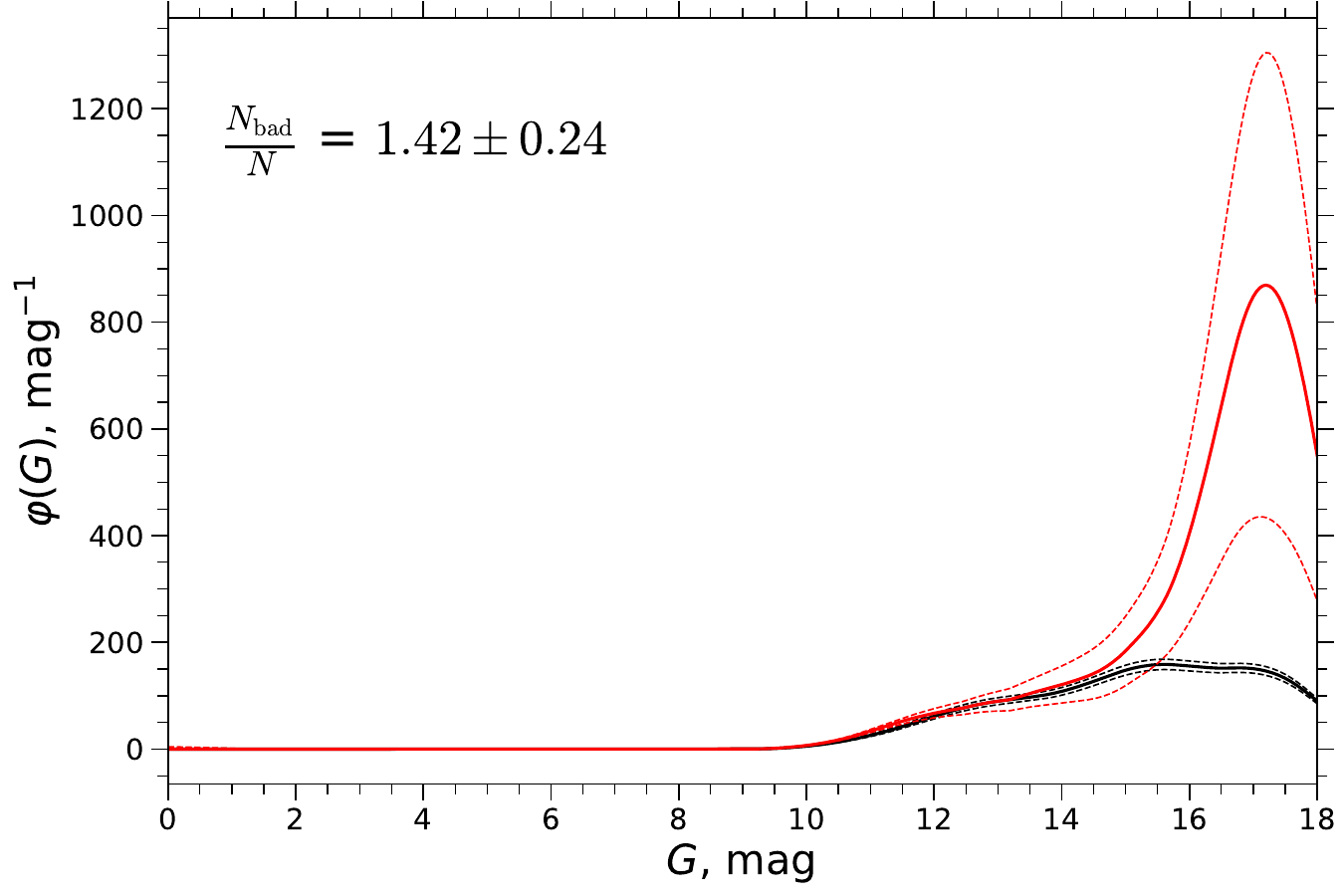}
	\caption{Brightness functions for NGC 2360. Solid lines are brightness functions, dashed lines are confidence intervals of $2\sigma$ width.
		Black lines are for stars from the Hunt\&Reffert sample \cite{H&R2024}, red lines --- with an addition of stars with poor astrometric solutions. }
	\label{Fi_G}
\end{figure}

The brightness function of the cluster NGC 2360 plotted by the stars of the Hunt\&Reffert sample \cite{H&R2024} is presented in Figure \ref{Fi_G} (black curve).
We plotted it using the one-dimensional KDE method \cite{Silverman1986,Merritt&Tremblay1994,KDE_OSC} with a biquadratic kernel.
We took the kernel half-width of 1.5 mag when constructing the brightness function for all clusters.
To account for stars with poor astrometric solutions, we need to plot a brightness function for such stars in a circle with radius $R_\textrm{b}$ and subtract the brightness function of field stars from it.
We used a ring with an inner radius of $R_\textrm{b}$ and an outer radius of $R_\textrm{b}\sqrt{2}$ as a comparison region (field stars).
If, as a result of subtraction, any values in the resulting brightness function were negative, we set these values equal to zero.
We obtain the overall brightness function by adding the brightness function from stars with poor astrometric solutions to the cluster brightness function by the \cite{H&R2024} sample.
As an example, the figure \ref{Fi_G} shows the overall brightness function for the cluster NGC 2360 (red curve).
The dotted lines show the confidence intervals of $2\sigma$ width plotted using the smoothed bootstrap method \cite{Seleznev2016}.

\begin{figure}[htbp]
	\includegraphics[width=1.0\textwidth]{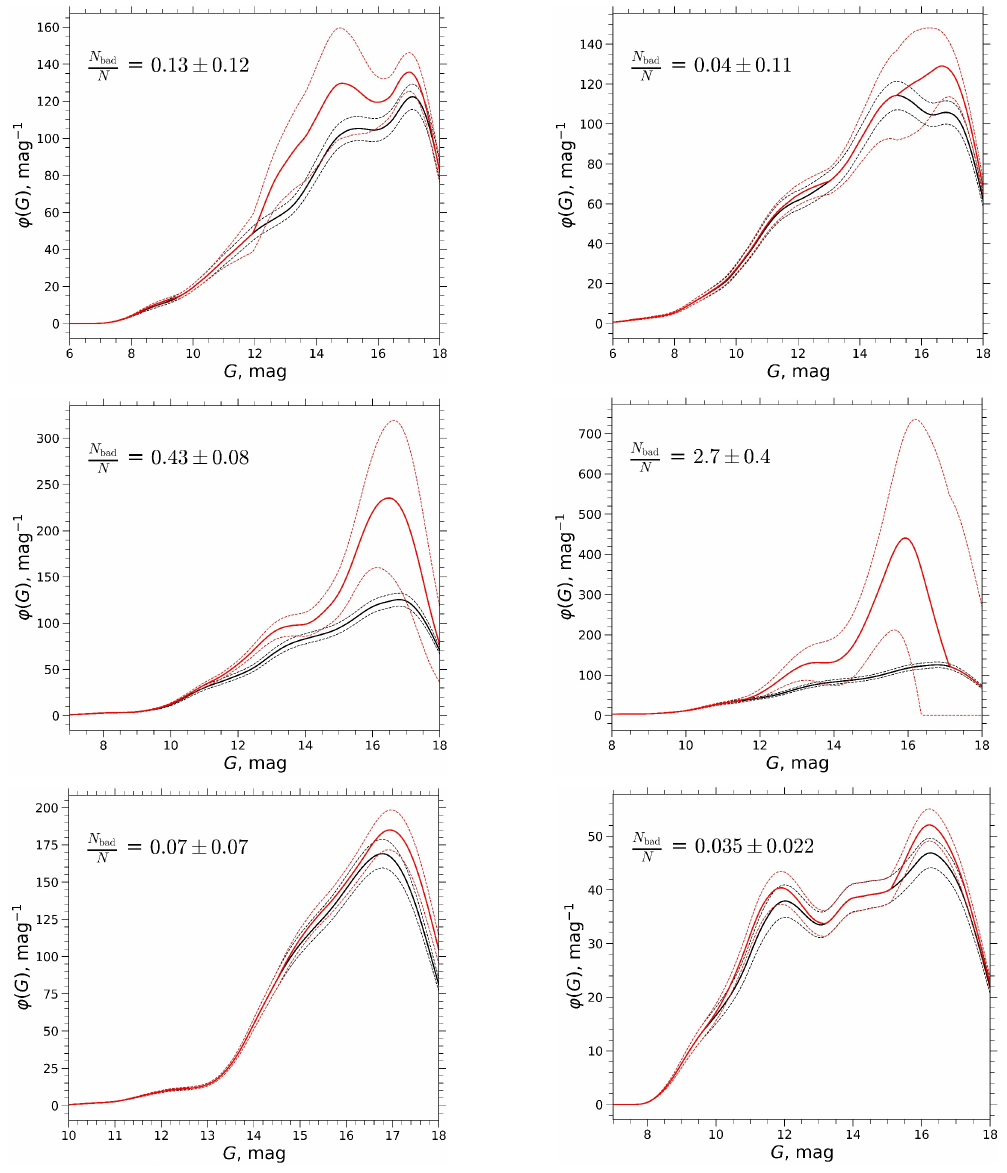}
	\caption{Brightness functions for clusters, from left to right, top to bottom: NGC 1093, NGC 2287, NGC 2301 (for the case $R_\textrm{c}=52.6\pm1.9$ arcmin), NGC 2301 (for the case $R_\textrm{c}=120\pm18$ arcmin), NGC 2420, NGC 2527.
		Designations are the same as in Fig. \ref{Fi_G}. }
	\label{LF1}
\end{figure}

\begin{figure}[htbp]
	\includegraphics[width=1.0\textwidth]{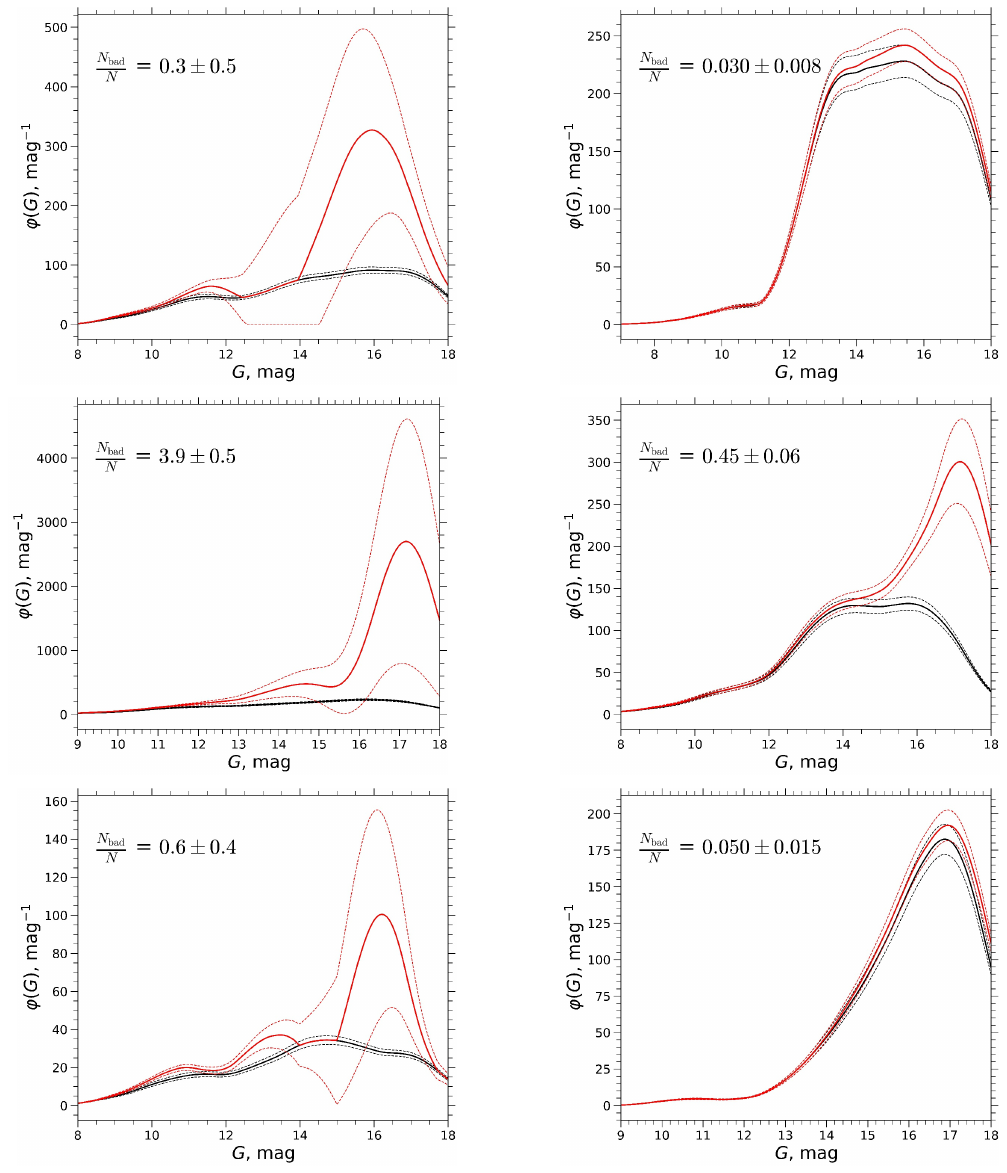}
	\caption{Brightness functions for clusters, left to right, top to bottom: NGC 2548, NGC 2682, NGC 3114, NGC 3766, NGC 5460, NGC 6649.
		Designations are the same as in Fig. \ref{Fi_G}. }
	\label{LF2}
\end{figure}

The brightness functions for the remaining clusters are presented in figures \ref{LF1}, \ref{LF2}.
In the upper left corner of these figures we show the relative number of possible cluster members with poor astrometric solutions of Gaia DR3.
We do not present the brightness function for the cluster NGC 188, since we do not find possible members of this cluster with poor astrometric solutions (see above).
It is evident that the brightness functions differ significantly in the region of dim stars $G\in[14,18]$ mag, just as in the case of NGC 3532 \cite{Tagaev}.
The likely reason is that the errors in the astrometric parameters of Gaia DR3 increase sharply in this range.
A large contribution to the number of such stars comes from unresolved binary and multiple systems near the Gaia resolution limit.
From the analysis of the brightness functions, we can conclude that the lack of stars with poor astrometric solutions has little effect on the overall brightness function, if a relative number of possible cluster members with poor astrometric solutions of Gaia DR3 $N_\textrm{bad}/N\lesssim0.15$.
For example, for the cluster NGC 2682, the overall brightness function and the brightness function for the stars of the Hunt\&Reffert sample \cite{H&R2024} coincide within the confidence intervals (figure \ref{LF2}).

\section{CONCLUSIONS}

In this paper, we have counted stars with poor astrometric solutions of Gaia DR3 in the regions of open star clusters: NGC 188, NGC 1039, NGC 2287, NGC 2301, NGC 2360, NGC 2420, NGC 2527, NGC 2548, NGC 2682 (M 67), NGC 3114, NGC 3766, NGC 5460, NGC 6649.
Our goal was to determine how many stars with poor solutions might be members of clusters.
The main selection criterion for such stars is that they fall within the region occupied by probable members of the cluster \cite{H&R2024} on the CMD.
For this purpose, we plot the Hess diagrams of the studied clusters.

In the wide vicinity of the clusters, we selected stars with $G\leq18$ mag, having only two-parameter solutions, or stars with 5- and 6-parameter solutions, but with the parameter RUWE>1.4, and/or a large value of the relative parallax error $\delta_\varpi/\varpi>0.2$.
Of these stars, we took only those that fell within the region occupied by the cluster on the Hess diagram.

The plotted radial surface density profiles for selected stars (with poor astrometric solutions) show a concentration towards the cluster center, except for the cluster NGC 188.
This suggests that some of these stars may be possible cluster members (they were missed by the traditional method of selecting of the probable cluster members).
The presence of a concentration towards the cluster center allows us to use the method of \cite{Seleznev2016} to determine the radius and number of stars in a cluster based on stars with poor astrometric solutions.
The obtained values of cluster radii for stars with poor solutions and the relative number of such stars are listed in the table \ref{table}.
The median mean relative number of stars with poor astrometric solutions is approximately 30\%.
This means that with the traditional method of identifying probable cluster members, on average 23\% of stars are lost because they have poor astrometric solutions.
Among the missing cluster members there may be a large number of unresolved binary and multiple systems with the component separations close to the Gaia resolution limit, since just for these objects the astrometric parameters either are determined unreliably or not determined at all.

We showed that at small values of the galactic latitude $b$ there are clusters with both large and small values of the relative number of stars with poor astrometric solutions.
As $b$ increases, the number of clusters with a large value of the relative number of stars with poor astrometric solutions decreases.
We can draw an upper envelope line that decreases with increasing of $b$.
There is also an increase of the upper envelope line of the relative number of stars with poor astrometric solutions with an increase in the average density of all stars in the studied region $\overline{F}$.
Then, we can assume that one of the sources of the poor astrometric solutions of Gaia DR3 is the high density of stars in the cluster region.

When taking into account stars with poor astrometric solutions of Gaia DR3, the cluster brightness function differs significantly from the brightness function plotted by stars of the Hunt\&Reffert sample \cite{H&R2024}, for $G\in[14,18]$ mag for clusters with the relative number of possible cluster members $N_\textrm{bad}/N\gtrsim0.15$.
In general, stars with magnitudes $G>14$ have poor astrometric solutions.
Differences in the brightness function should lead to the differences in the mass spectrum.
Consequently, the taking into account stars with poor astrometric solutions increases the photometric mass of the cluster.
Taking into account stars with poor astrometric solutions will reduce the difference between photometric and virial mass.
It is also possible to correct the initial mass and luminosity functions \cite{Piskunov,Kroupa}.

\section*{FUNDING}
This work was supported by the Ministry of science and higher education of the Russian Federation by an agreement FEUZ-2023-0019.

\begin{acknowledgments}
This work has made use of data from the European Space Agency (ESA) mission Gaia (https://www.cosmos.esa.int/gaia), processed by the Gaia Data Processing and Analysis Consortium (DPAC, https://www.cosmos.esa.int/web/gaia/dpac/consortium).
Funding for the DPAC has been provided by national institutions, in particular the institutions participating in the Gaia Multilateral Agreement.
\end{acknowledgments}

\section*{CONFLICT OF INTEREST}

The authors declare no conflict of interest.


\bibliographystyle{maik}
\bibliography{Tagaev}

\end{document}